\documentclass{PoS}
\usepackage{graphicx}
\usepackage{cite}
\newcommand{\ntsj}[2]{#1\,{}^{3\!}S_{#2}}
\newcommand{\ntdj}[2]{#1\,{}^{3\!}D_{#2}}

\title{Exes and why Z?\\
Some charming and beautiful observations.}
\ShortTitle{Exes and why Z?}
\author{\speaker{Eef van Beveren}\\
Centro de F\'{\i}sica Computacional,
Departamento de F\'{\i}sica,
Universidade de Coimbra, P-3004-516 Coimbra, Portugal\\
E-mail: \email{eef@teor.fis.uc.pt}}
\author{George Rupp\\
Centro de F\'{\i}sica das Interac\c{c}\~{o}es Fundamentais,
Instituto Superior T\'{e}cnico,
Universidade T\'{e}cnica de Lisboa, Edif\'{\i}cio Ci\^{e}ncia,
P-1049-001 Lisboa, Portugal\\
E-mail: \email{george@ist.utl.pt}}
\abstract{Threshold enhancements like the $X(4660)$ and depletion effects
as the $X(4260)$ are listed as $c\bar{c}$ resonances in the Particle Data
Group tables. We will discuss these observations, and present
a list of further $c\bar{c}$ enhancements,
which are more likely to represent true vector charmonium excitations.\\
We will furthermore discuss
the importance of the observed $Z$ resonances, viz.\
$Z(4050)$, $Z(4250)$, and $Z(4430)$,
for the family of charm-strange mesons.\\
Another piece of very important information
that can be extracted from the present data is
the universal, flavor independent frequency of 190 MeV for mesons,
due to the quark-antiquark oscillations within the glue environment.\\
Finally, we will show hints from the data
at a further flavor-independent quantity, having a value of 76$\pm$2 MeV,
the origin of which is not yet understood.}
\FullConference{The Xth Nicola Cabibbo International Conference
on Heavy Quarks and Leptons,\\ October 11--15, 2010\\ Frascati (Rome) Italy}
\begin{document}
In the recent past, we have developed a powerful formalism
that does not necessarily rely on a specific choice
for the dynamics of the quark-antiquark system,
and which may reproduce
the observed cross sections in meson-meson scattering
\cite{AP324p1620}
and in electron-positron annihilation reactions
\cite{AP323p1215}.
It is based on the assumption that observed structures in
non-exotic meson-meson scattering and production
are dominated by quark-antiquark ($q\bar{q}$) resonances.
In the Resonance-Spectrum Expansion (RSE) \cite{IJTPGTNO11p179},
one may, in principle, ignore any specific dynamics
of the $q\bar{q}$ system, since the RSE expressions only contain
the resulting $q\bar{q}$ spectrum as input. Here, we will
concentrate on the harmonic-oscillator approximation of the RSE (HORSE),
for which the spectrum is equidistant with a level spacing of 380 MeV,
independent of flavor.

It has almost become an automatism for experimental collaborations
to interpret observed enhancements in scattering and production cross
sections as resonances. On the other hand, most theorists compare
the central masses of such enhancements to all sorts of model
calculations, without even worrying how to reproduce the observed
enhancements themselves. However, we have shown at different occasions
that there exist at least four different types of enhancements, two of
which are nonresonant and therefore should not be described by any kind
of quark models that disregard decay thresholds. A careful study of
both meson-meson data and the implications of our theory for scattering
and production reveals that there exist enhancements which can be
described by either {\it genuine} \/or {\it accidental resonances},
besides {\it threshold enhancements} \/and {\it depletion effects}.
\vspace{5pt}

{\bf Genuine resonances} manifest themselves as poles
in the scattering and production amplitudes,
which are one-to-one related to $q\bar{q}$ states,
or to possible other quark and/or gluonic configurations.
This can be studied by gradually turning off the coupling between
the confined $q\bar{q}$ channel(s) and the meson-meson channels
in the scattering/production amplitudes.
In this process, such poles end up at the positions corresponding to
the real energy levels of the confinement states.
\cite{PRD21p772,PRD27p1527}.
\vspace{5pt}

{\bf Accidental resonances}
are also designated by ``dynamically generated resonances'' in modern
literature. They are generated solely by the coupling between $q\bar{q}$
and meson-meson channels. In the process of turning off the coupling,
such poles disappear into the continuum, with ever increasing negative
imaginary parts, thus becoming resonances with infinite widths
\cite{ZPC30p615}.
\vspace{5pt}

{\bf Threshold enhancements}
occur in electron-positron annihilation reactions
at the opening of new channels.
In some cases these enhancements are large
and allow for the discovery of
genuine or accidental resonances in their tails
\cite{EPL85p61002},
or even on top of the peaking structures
\cite{PRD80p074001}.
Threshold enhancements themselves
are nonresonating and do not correspond to poles
in the scattering/production amplitudes.
\vspace{5pt}

{\bf Depletion effects}
may resemble somewhat enhancements
but stem from a process of competition between decay channels,
whereby genuine or accidental resonances as well as
threshold enhancements in one channel
deplete the signal in another channel
\cite{PRL105p102001}.
In the channel of depleted signal,
resonances and threshold enhancements
are observed by dips instead of bumps in the cross sections,
in contrast with what is observed in the other channel.
The remaining structure, with the dips,
may be be mistaken for a number of resonances between the dips,
in many experimental and theoretical analyses.
\vspace{5pt}

In most approaches to strong interactions as well as
light- and heavy-quark physics, the latter two types of enhancements
cannot be described, as one is compelled to conclude from the countless
model calculations outlined in Ref.~\cite{ARXIV10105827}, which,
without exception, completely ignore the possibility of nonresonant
enhancements.

The real $q\bar{q}$ resonances are often quite modest enhancements,
and must be searched for with great care in experimental data.
Recently, we have identified a few candidates for new vector charmonium states
\cite{EPL85p61002,PRL105p102001,ARXIV09044351,ARXIV10044368,ARXIV10053490}.
In Table~\ref{ccbar} we compare these findings
to the predictions from the pure HO spectrum
(first column, ``HORSE quenched''),
which are given by
\begin{equation}
E_{q,n\ell}=2m_{q}+\omega\left( 2n+\ell +\frac{3}{2}\right)
\;\;\; ,
\label{quenched}
\end{equation}
for $q=c$, with the charm quark mass $m_c=1.562$ GeV
and oscillator frequency $\omega =0.190$ GeV
taken from Ref.~\cite{PRD27p1527}.
\begin{table}[htbp]
\begin{center}
\begin{tabular}{|c|ll|}
\hline
HORSE quenched & $\psi (D)$ & $\psi (S)$\\
\hline
3.789 & 3.773 (1D \cite{JPG37p075021}) & 3.686 (2S \cite{JPG37p075021})\\
4.169 & 4.153 (2D \cite{JPG37p075021}) & 4.039 (3S \cite{JPG37p075021})\\
4.549 & $\approx$4.56 (3D \cite{ARXIV09044351,PRL105p102001})
& 4.421 (4S \cite{JPG37p075021}) \\
4.929 & $\approx$4.89 (4D \cite{EPL85p61002,ARXIV10053490})
& $\approx$4.81 (5S \cite{EPL85p61002,ARXIV10053490})\\
5.309 & $\approx$5.29 (5D \cite{ARXIV09044351})
& $\approx$5.13 (6S \cite{ARXIV09044351})\\
5.689 & $\approx$5.66 (6D \cite{ARXIV10044368})
&  $\approx$5.44  (7S \cite{ARXIV10044368})\\
6.069 &  -- (7D) & $\approx$5.91 (8S \cite{ARXIV10044368})\\
\hline
\end{tabular}
\end{center}
\caption[]{\small
Central masses (GeV) of the higher vector charmonium states,
including the well-known ones for three decades \cite{JPG37p075021}
and those extracted by us from data.
}
\label{ccbar}
\end{table}
The HORSE quenched $nS$ and $(n\!-\!1)D$  $c\bar{c}$ masses
are degenerate.
We find that the bare $c\bar{c}$ states turn into
bound states below the $D\bar{D}$ threshold, or resonances thereabove,
by unquenching the $c\bar{c}$ states
through the insertion of  open-charm meson-meson loops
\cite{PRD21p772,IJTPGTNO11p179},
also for bound states below the $D\bar{D}$ threshold.
The $S$ states (third column of Table~\ref{ccbar})
have central masses of about 100--200 MeV below the unquenched levels,
whereas the $D$ states (second column of Table~\ref{ccbar})
undergo much smaller mass shifts. The exact values of
these mass shifts also depend on the specific positions
of the open-charm thresholds with respect to the quenched $c\bar{c}$
states.

Results for beautonium alias bottomonium, taking $m_{b}=4.724$ GeV
\cite{PRD27p1527} in Eq.~(\ref{quenched}) with $q=b$, are given in Table~\ref{bbbar}.
We observe a $b\bar{b}$ spectrum which is very similar
to the $c\bar{c}$ spectrum of Table~\ref{ccbar},
just shifted towards higher masses by about 6.3 GeV.
However, our particle assignments are somewhat
different from what one finds in most of the literature.

\begin{table}[htbp]
\begin{center}
\begin{tabular}{|c|ll|}
\hline
HORSE quenched & $\Upsilon (D)$ & $\Upsilon (S)$\\
\hline
10.113 & 10.098 (1D \cite{ARXIV10094097}) & 10.023 (2S \cite{JPG37p075021})\\
10.493 & 10.495 (2D \cite{ARXIV10094097}) & 10.355 (3S \cite{JPG37p075021})\\
10.873 & 10.865 (3D \cite{JPG37p075021}) & 10.735 (4S \cite{ARXIV09100967}) \\
11.253 &  -- (4D) & 11.019 (5S \cite{JPG37p075021})\\
\hline
\end{tabular}
\end{center}
\caption[]{\small
Energy levels (GeV) of the HORSE quench\-ed $b\bar{b}$ spectrum;
bound-state and central resonance masses (GeV)
as deduced from experiment for the $\Upsilon$ vector states.}
\label{bbbar}
\end{table}

The experimental identification of the resonance
at 10.845 GeV (CUSB) or 10.868 GeV (CLEO)
and the resonance
at 11.02 GeV (CUSB) or 11.019 GeV (CLEO)
with the $\Upsilon (5S)$ and $\Upsilon (6S)$, respectively,
was apparently inspired by the corresponding model predictions
of Godfrey and Isgur \cite{PRD32p189}, i.e., at 10.88 GeV and 11.10 GeV,
respectively. However, we rather identify these resonances with
the $\Upsilon (3D)$ and $\Upsilon (5S)$ states, respectively,
on the basis of the level schemes in
Tables~\ref{ccbar} and \ref{bbbar}
\cite{PRD21p772,PRD27p1527}.
\vspace{10pt}

{\bf $Z$ mesons}\\
Matsuki, Morii, and Sudoh \cite{PLB669p156}
were the first to suggest
that the observed $Z^{+}$ resonances might be higher excitations
of the $D_{s}^{\ast}$(2112) state.
Now, since one indeed finds for the central masses
of the $Z^{+}$(4430) (alias $X(4430)^\pm$ \cite{JPG37p075021})
and $Z^{+}$(4050) (alias $X(4050)^\pm$ \cite{JPG37p075021})
a mass difference of about 380 MeV,
it is interesting to check the plausibility of the
suggestion in Ref.~\cite{PLB669p156}.
\begin{figure}[htbp]
\centering
\begin{tabular}{|c|c|c|}
\hline & & \\ [-11pt]
\includegraphics[width=130pt]{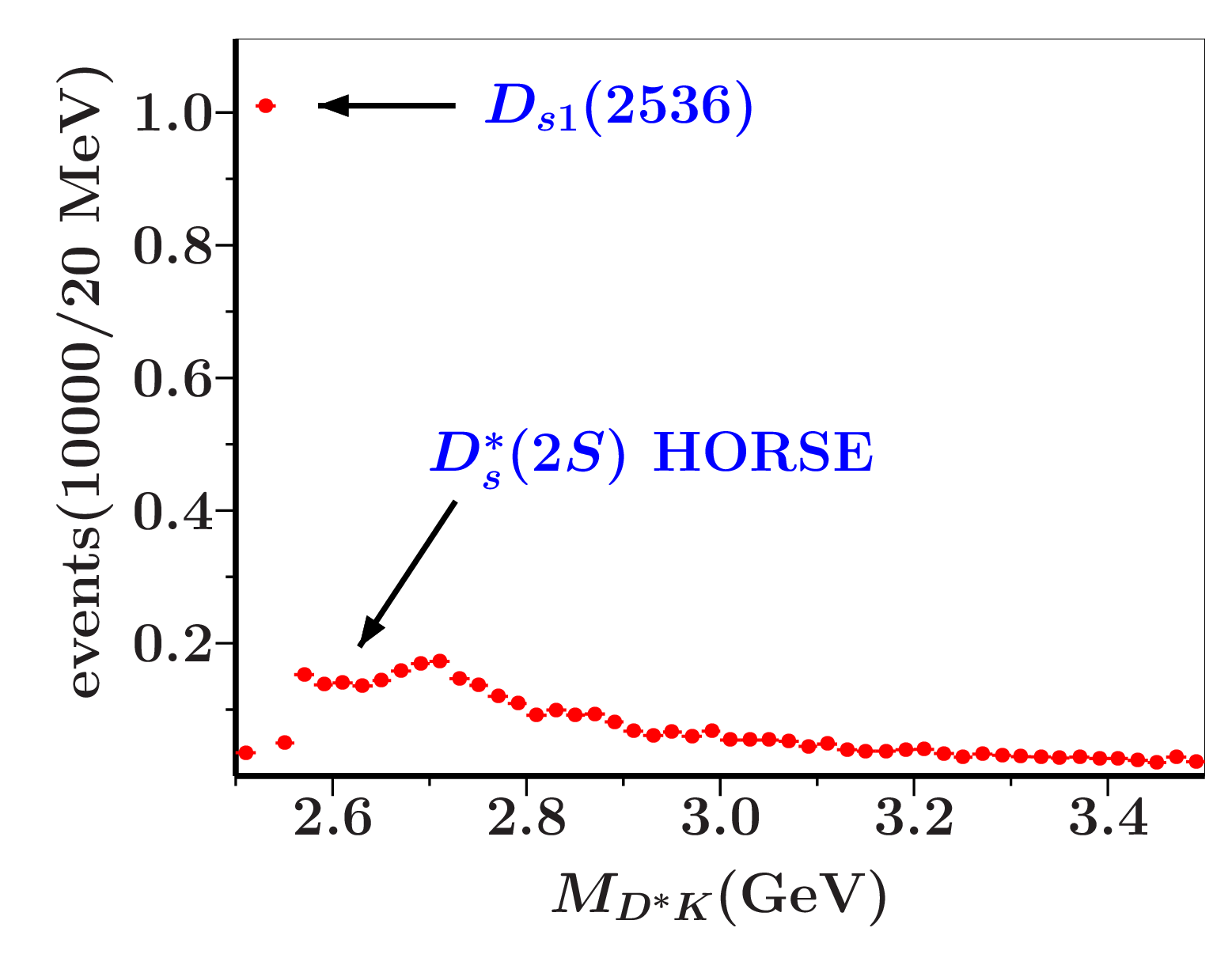} &
\includegraphics[width=130pt]{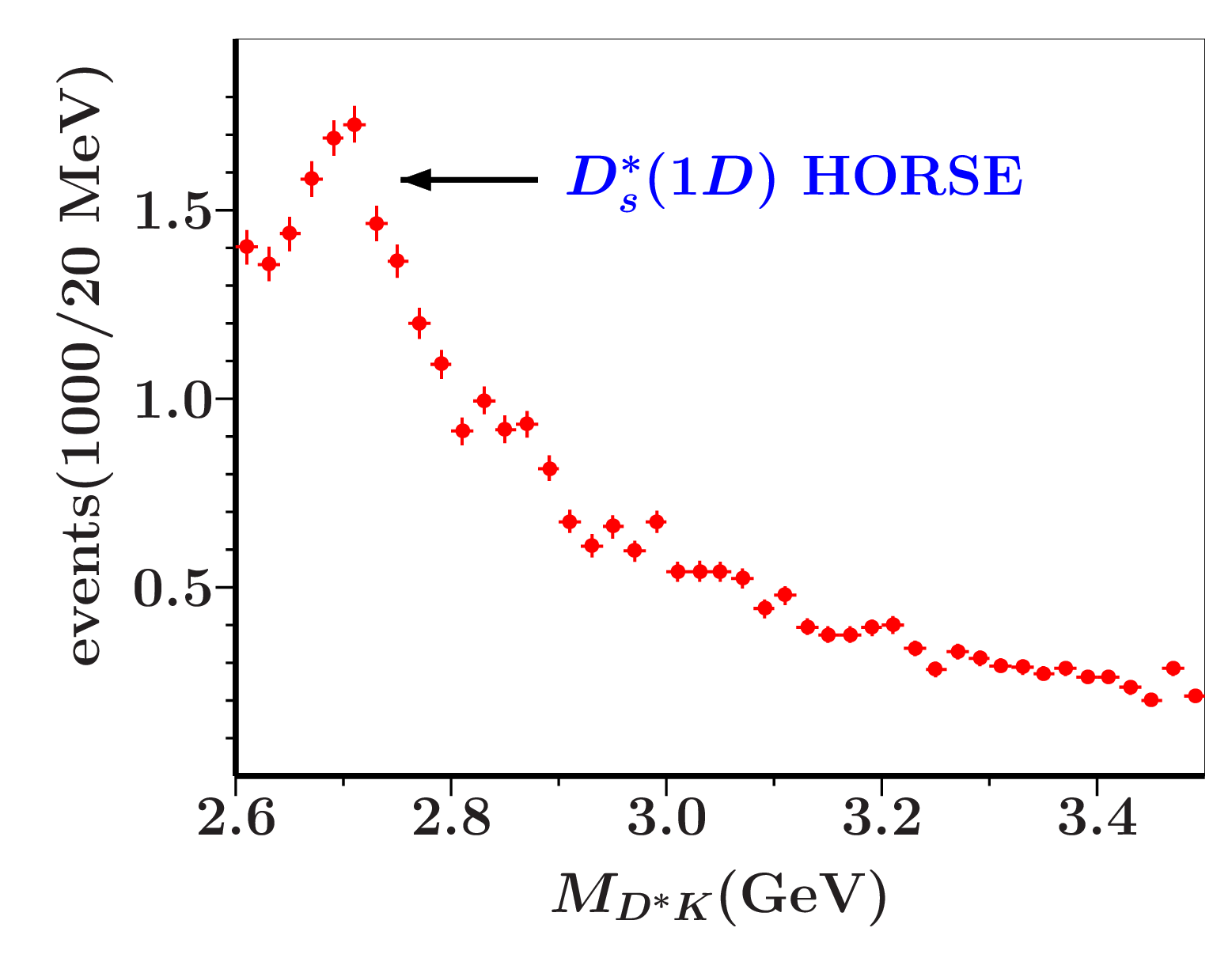} &
\includegraphics[width=130pt]{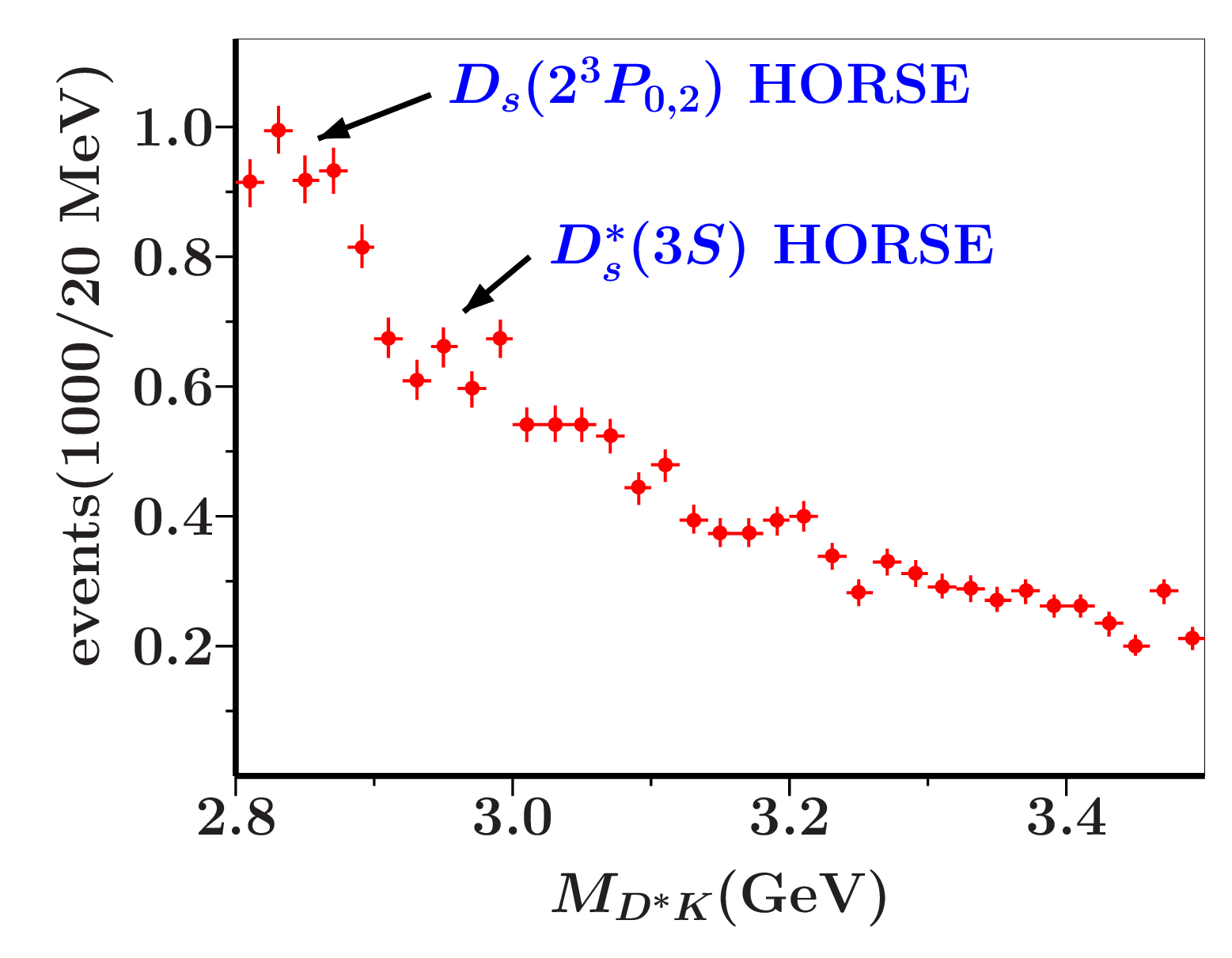} \\
\hline & & \\ [-11pt]
\includegraphics[width=130pt]{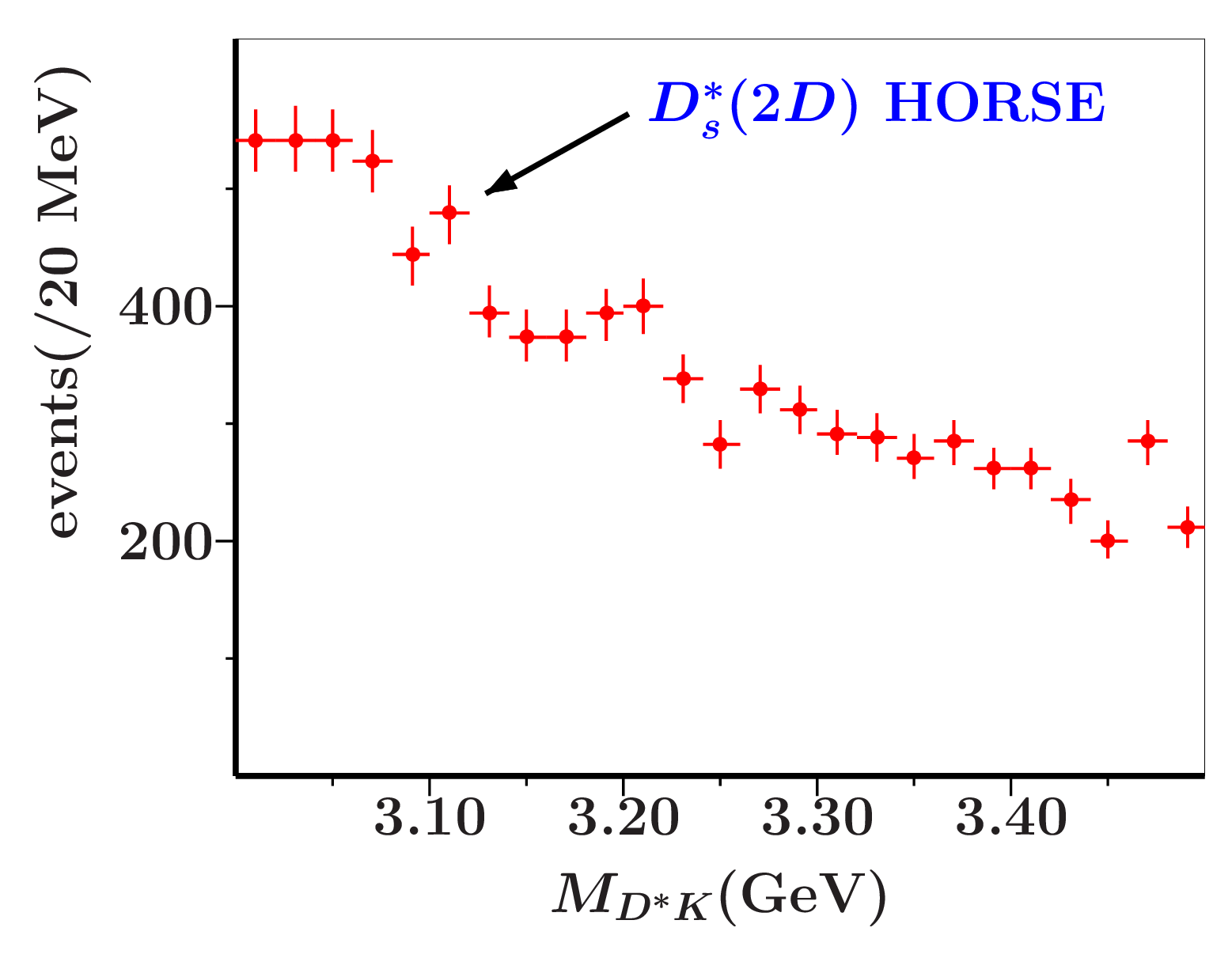} &
\includegraphics[width=130pt]{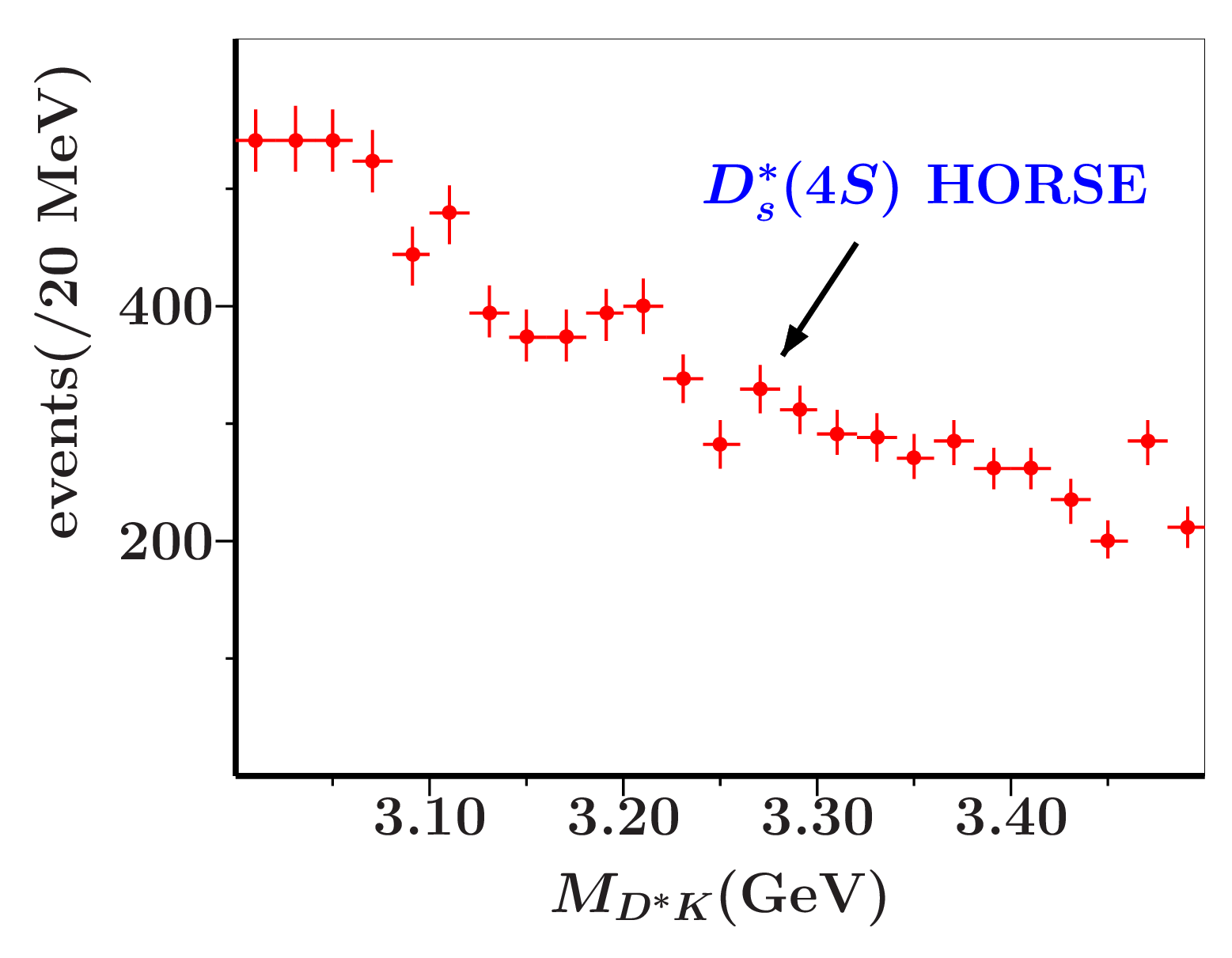} &
\includegraphics[width=130pt]{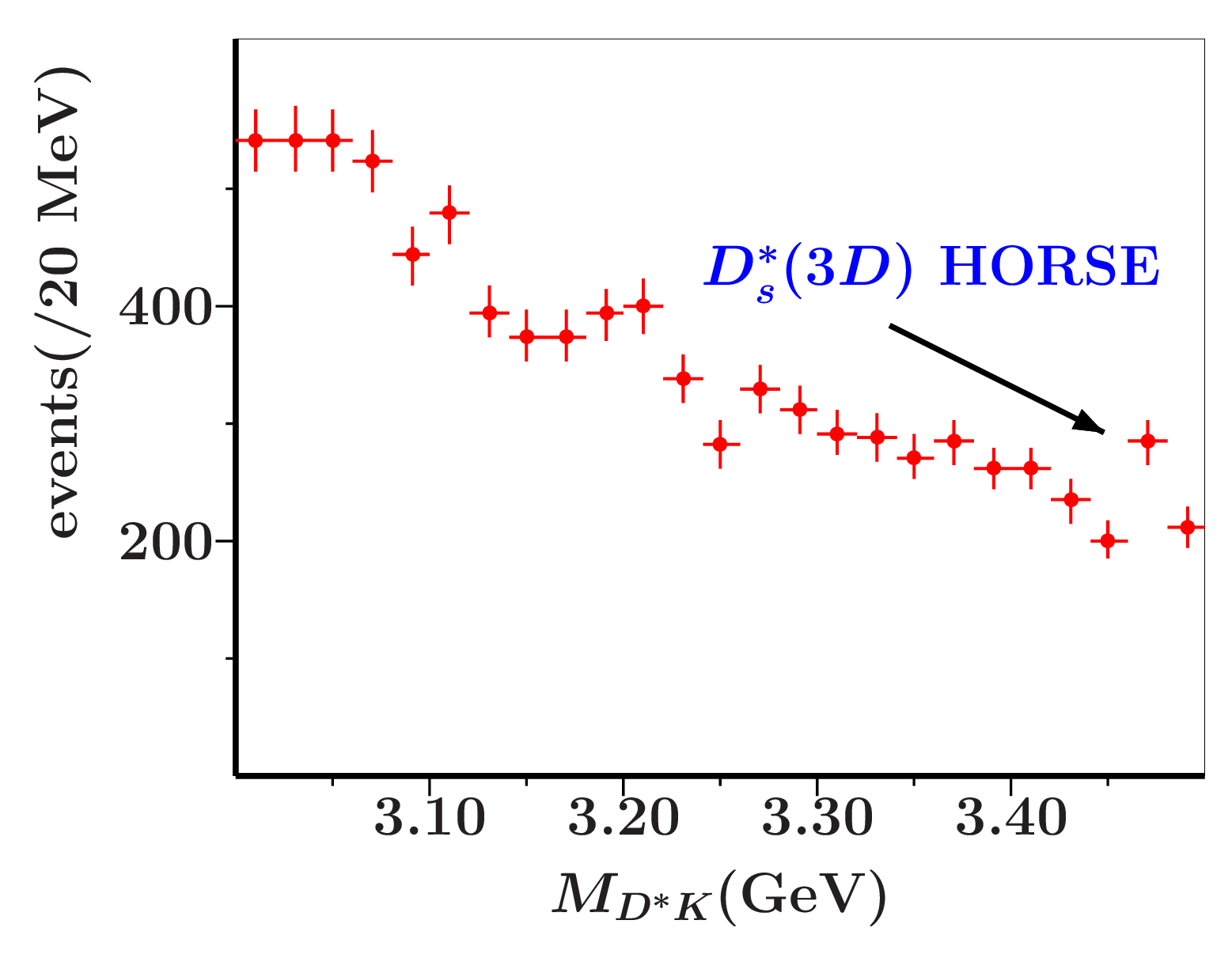} \\
\hline
\end{tabular}
\caption[]{\small
$D^{\ast}K$ invariant-mass distribution from BABAR \cite{PRD80p092003}
and indications of possible $D_{s}^{\ast}$ excitations.
}
\label{DstarK}
\end{figure}
If one takes a charm quark mass $m_c=1.562$ GeV
and a strange quark mass $m_s=0.508$ GeV, as in
Ref.~\cite{PRD27p1527},
one obtains from Eq.~(\ref{quenched}) for
the degenerate $(\ntdj{5}{1}$, $\ntsj{6}{1})$ pair
a mass of 4.255 GeV,
and for the degenerate $(\ntdj{6}{1}$, $\ntsj{7}{1})$ pair
a mass of 4.635 GeV. This suggests
that the $Z^{+}$(4250) (alias $X(4250)^\pm$ \cite{JPG37p075021})
could be the $D_{s}^{\ast}\left(\ntdj{5}{1}\right)$.
The $Z^{+}$(4050) and the $Z^{+}$(4430),
with a mass shift of roughly 200 MeV due to meson loops,
may then be identified with the
$D_{s}^{\ast}\left(\ntsj{6}{1}\right)$
and $D_{s}^{\ast}\left(\ntsj{7}{1}\right)$, respectively.

Fortunately, we have at our disposal data on $D^{\ast}K$
from BABAR \cite{PRD80p092003},
which is the channel where one expects to observe
the higher excitations of the $D_{s}^{\ast}$(2112).
These data are displayed in Fig.~\ref{DstarK}.
In the various plots here one can see structures in the data
at approximately the energies for which the HORSE predicts
$D_{s}^{\ast}$ excitations.
Consequently, the above suggestion is very well possible.
We summarize our observations in Table~\ref{Zees}.
\begin{table}[htbp]
\begin{center}
\begin{tabular}{|l|c|l|}
\hline
states & HORSE quenched & observations\\
\hline
$1S$ & 2.355 & $D_{s}^{\ast}$(2.112) \cite{JPG37p075021}\\
$2S$, $1D$ & 2.735 & $2S$ difficult, $1D$: $D_{s}^{\ast}$(2.71)\\
$3S$, $2D$ & 3.115 & $3S$ and $2D$ both possible\\
$4S$, $3D$ & 3.495 & $4S$ and $3D$ both possible\\
$5S$, $4D$ & 3.875 & --\\
$6S$, $5D$ & 4.255 & $Z^{+}$(4050) and $Z^{+}$(4250)\\
$7S$, $6D$ & 4.635 & $Z^{+}$(4430), $6D$ too high\\
\hline
\end{tabular}
\end{center}
\caption[]{\small
Energy levels (GeV) of the HORSE quench\-ed $c\bar{s}$ spectrum;
possible interpretation in terms of excited $D_s^\ast$ or $Z^+$ resonances.}
\label{Zees}
\end{table}
With respect to our $1D$ assignment for the $D_{s}^{\ast}$(2.71),
one must note the following.
The $2S$ assignment by BABAR stems solely from branching ratios
determined in Ref.~\cite{PRD77p014012}. If we calculate these branching
ratios, using the coupling constants of Ref.~\cite{ZPC21p291},
we find a ratio that is more than 3 times too large for the $2S$ state,
and 77\% of the experimental result for the $1D$.
Moreover, since $S$ and $D$ states get mixed by meson loops,
it is more likely to assume that the $D_{s}^{\ast}$(2.71) is mainly $1D$.
Unfortunately, the $D_{s}^{\ast}\left(\ntsj{2}{1}\right)$
resonance is in the HORSE expected to be not far from the $D_{s1}$(2536),
and since the latter state produces a huge enhancement in the data,
it may take quite some effort to find the
$D_{s}^{\ast}\left(\ntsj{2}{1}\right)$ in its vicinity.
\vspace{10pt}

{\bf A flavor-independent interference effect}\\
In Refs.~\cite{PRD79p111501R,ARXIV10095191}
we described an interference effect that is visible
in the cross sections of different annihilation processes.
Although we have no explanation so far for such a phenomenon,
it might add a second constant to the list
of universal parameters for mesons.
\begin{figure}[htbp]
\centering
\includegraphics[width=230pt]{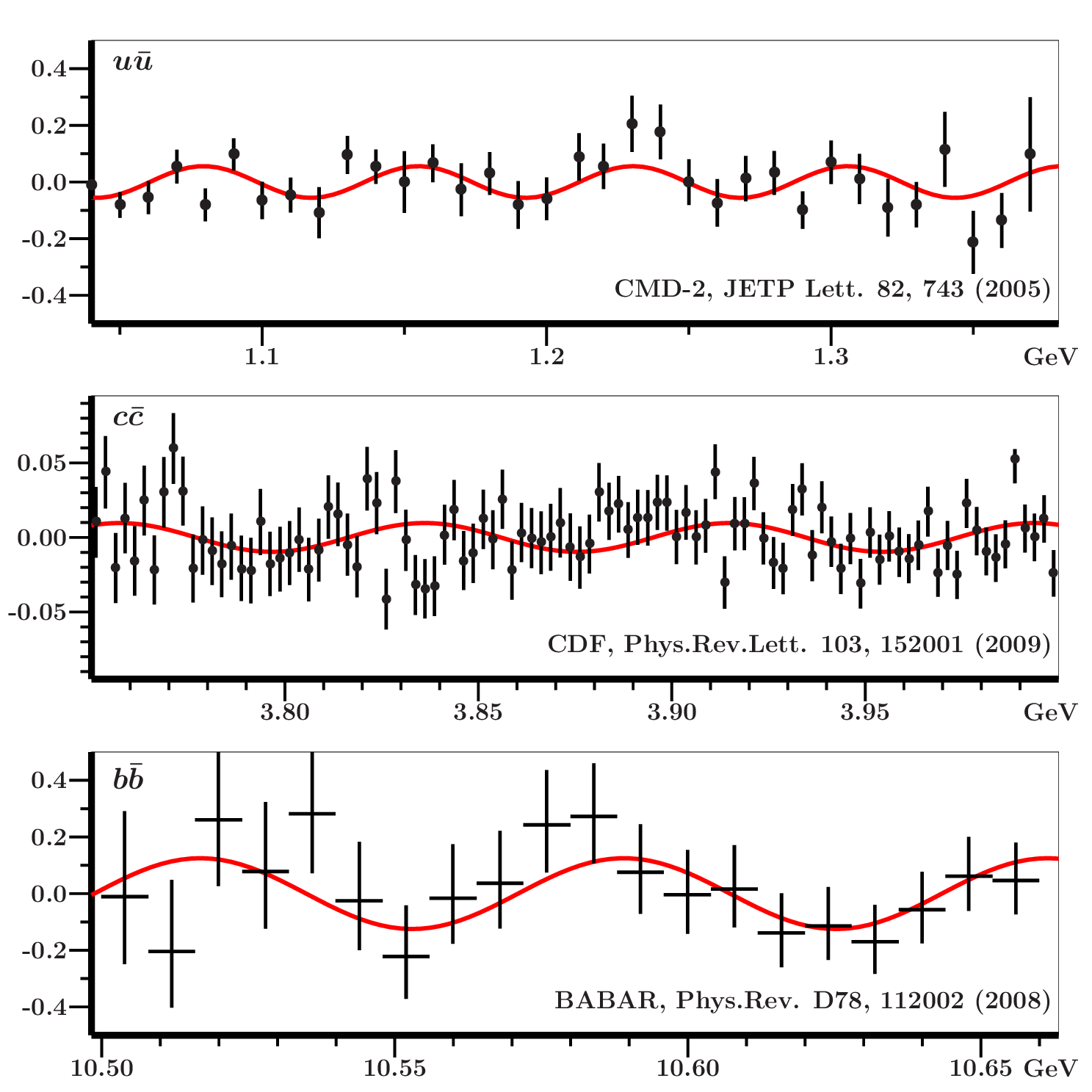}
\caption[]{\small
A flavor-independent interference effect is observed by us
in data corresponding to processes for light quarks, $c\bar{c}$, and
$b\bar{b}$ \cite{ARXIV10095191}.
}
\label{interference}
\end{figure}
\vspace{10pt}

{\bf Summarizing},
we have argued that a constant radial level splitting
of about 380 MeV is consistent with light- and heavy-meson spectra.
Furthermore, we presented a possible second flavor-independent observable
for mesons.
We also hinted at the possibility that the mysterious $Z^{+}$ resonances
are just higher excitations in the $D_{s}^{\ast}$ spectrum.
Finally, we discussed ex-resonances that are
either threshold enhancements or leftovers due to depletion effects.
\vspace{10pt}

One of us (EvB) thanks the organizers
of {\it The Xth International Conference on Heavy Quarks and Leptons}
\/for arranging a very pleasant
setting for discussion and exchange of ideas.
This work was supported in part by the {\it Funda\c{c}\~{a}o para a
Ci\^{e}ncia e a Tecnologia} \/of the {\it Minist\'{e}rio da Ci\^{e}ncia,
Tecnologia e Ensino Superior} \/of Portugal, under contract
CERN/\-FP/\-109307/\-2009.

\newcommand{\pubprt}[4]{#1 {\bf #2}, #3 (#4)}
\newcommand{\ertbid}[4]{[Erratum-ibid.~#1 {\bf #2}, #3 (#4)]}
\def\AP{Ann.\ Phys.}
\def\EPL{Europhys.\ Lett.}
\def\IJTPGTNO{Int.\ J.\ Theor.\ Phys.\ Group Theor.\ Nonlin.\ Opt.}
\def\JPG{J.\ Phys.\ G}
\def\PLB{Phys.\ Lett.\ B}
\def\PRD{Phys.\ Rev.\ D}
\def\PRL{Phys.\ Rev.\ Lett.}
\def\ZPC{Z.\ Phys.\ C}

\end{document}